\documentclass[]{aa}
\usepackage{graphicx}
\usepackage{psfig}
\usepackage{amssymb}
\usepackage{amsmath}
\usepackage{verbatim}
\usepackage{natbib}
\usepackage{rotate}
\usepackage{lscape}
\usepackage{aalongtable}
\usepackage{supertabular}
\usepackage{mathbbol}
\usepackage{bm}
\usepackage{mathtools}
\usepackage{enumerate}
\usepackage{booktabs}

\bibpunct{(}{)}{;}{a}{}{,}

\newcommand{\appropto}{\mathrel{\vcenter{
  \offinterlineskip\halign{\hfil$##$\cr
    \propto\cr\noalign{\kern2pt}\sim\cr\noalign{\kern-2pt}}}}}

\def\mathY{\bm{\mathcal{Y}}}
\def\Bchi{\large\bm{\mathcal{\chi}}}
\def\PVec{\textbf{x}}

\def\MR{\textbf{R}_k}
\def\MS{\textbf{S}_k}
\def\MW{\textbf{W}}

\begin{document}      
	      

\title{Non-linear Kalman filters for calibration in radio interferometry}

\author{C. Tasse\inst{1,2,3}}

\institute{
GEPI, Observatoire de Paris, CNRS, Universit\'e Paris Diderot,
5 place Jules Janssen, 92190 Meudon, France
\and
SKA South Africa, 3rd Floor, The Park, Park Road, Pinelands, 7405, South Africa
\and
Department of Physics \& Electronics, Rhodes University, PO Box 94,
Grahamstown, 6140, South Africa}
\date{Received <date> / Accepted <date>}

\abstract{The data produced by the new generation of interferometers are
affected by a large variety of partially unknown complex effects such
as pointing errors, phased array beams, ionosphere, troposphere,
Faraday rotation, or clock drifts.  Most algorithms addressing
direction-dependent calibration solve for the effective Jones
matrices, and cannot constrain the underlying physical quantities of
the Radio Interferometry Measurement Equation (RIME).  A related
difficulty is that they lack robustness in the presence of low
signal-to-noise ratios, and when solving for moderate to large number
of parameters they can be subject to ill-conditioning. Those effects
can have dramatic consequences in the image plane such as source or
even thermal noise suppression. The advantage of solvers directly
estimating the physical terms appearing in the RIME, is that they can
potentially reduce the number of free parameters by orders of
magnitudes while dramatically increasing the size of usable data,
thereby improving conditioning.

We present here a new calibration scheme based on a non-linear version
of Kalman filter that aims at estimating the physical terms appearing
in the RIME.
We enrich the filter's structure with
a tunable data representation model, together with an augmented
measurement model for regularization. We show using simulations that
it can properly estimate the physical effects appearing in the
RIME.
We found that this approach is
particularly useful in the most extreme cases such as when ionospheric
and clock effects are simultaneously present. Combined with the
ability to provide prior knowledge on the expected structure of the
physical instrumental effects (expected physical state and dynamics),
we obtain a fairly cheap algorithm that we believe to be
robust, especially in low signal-to-noise regime. Potentially the use of filters and other similar methods can
represent an improvement for calibration in radio interferometry,
under the condition that the effects corrupting visibilities are
understood and analytically stable. Recursive algorithms are
particularly well adapted for pre-calibration and sky model estimate
in a streaming way. This may be useful for the SKA-type instruments
that produce huge amounts of data that have to be calibrated before
being averaged.

}

\authorrunning{C. Tasse}

\titlerunning{Non-linear Kalman filters and regularisation techniques in radio interferometry}
   \maketitle


\section{Introduction}

The new generation of interferometers are characterized by very wide
fields of view, large fractional bandwidth, high sensitivity, and high
resolution. At low frequency (LOFAR, PAPER, MWA) the cross-correlation
between voltages from pairs of antenna (the visibilities) are affected
by severe complex baseline-time-frequency Direction Dependent Effects
(DDE) such as the complex phased array beams, the ionosphere and its
associated Faraday rotation, the station's clock drifts, and the sky
structure. At higher frequency, the interferometers using dishes are less affected by ionosphere, but troposphere, pointing errors and
dish deformation play an important role.

\def\X{\text{X}}
\def\Y{\text{Y}}
\def\u{u}
\def\v{v}
\def\w{w}
\def\l{l}
\def\m{m}
\def\n{n}

\subsection{Direction dependent effects and calibration issues}
\label{sec:PhysPeal}

A large variety of solvers have been developed to tackle the
direction-dependent calibration problems of radio interferometry. In this paper, for the
clarity of our discourse, we classify them in two categories. The
first and most widely used family of algorithms (later referred as the
{\it Jones-based Solvers}) aim at estimating the {\it apparent} net
product of various effects discussed above. The output solution is a
Jones matrix per time-frequency bin per antenna, per
direction
\citep[see][and references therein]{Yatawatta08,Noordam10}. Sometimes
the solutions are used to derive physical parameters \citep[for
  example][in the cases of ionosphere and beam shape
  respectively]{Intema09,Yatawatta13}. The second family of solvers
estimate directly from the data the physical terms mentioned above
that give rise to a set of visibility (later referred as the
{\it Continuous} or {\it Physics-based Solvers}). Such solvers are used for
direction-independent calibration in the context of
 fringe-fitting for VLBI \citep[][and references
  therein]{Cotton95,Walker99} to constrain the clock states and drifts
(also referred as delays and fringe rates). \citet{SB:pointing} and
\citet{Smirnov2011qmc} have presented solutions to the direction-dependent
calibration problem of pointing error. It is important to note that
deconvolution algorithms, are also Physics-based solvers estimating
the sky brightness, potentially taking DDE calibration solution into
account \citep[][]{Bhatnagar08,Bhatnagar12,Tasse13}. Latest imaging solvers can
also estimate spectral energy distribution parameters
\citep[][]{Rau11,Junklewitz14}. Most of these imaging algorithms are
now well understood in the framework of compressed sensing theory
\citep[see][for a review]{McEwen11}. Their goals,
constrains and methods are however very different from purely
calibration-related algorithms, and we will not discuss them further
in this paper.

Jones-based and Physics-based solvers have both advantages and
disadvantages. The main issue using Physics-based solvers is that the
system needs to be modeled accurately, while analytically complex
physics can intervene before measuring a given visibility. Jones-based
algorithms solving for the effective Jones matrices do not suffer from
this problem, because no assumptions have to be made about the physics underlying the building of a visibility
(apart from the sky model that is assumed).

However, one important disadvantage of Jones-based solvers over
Physics-based solvers for DDE calibration is that they lack robustness when solving for a
large number of parameters and can be subject to
ill-conditioning. 
This can have dramatic effects in the image plane,
such as source suppression. In the most extreme case, those algorithms
can artificially decrease noise in the calibrated residual maps by
over-fitting the data. This easily drives artificially high dynamic
range estimates. In fact, hundreds of parameters ({\it i.e. of
  directions}) per antenna, polarization, can
correspond to tens of thousands of free parameters per time and
frequency bin. 
The
measurement operator being highly non-linear, for given data set and
process space, it is often hard to know whether ill conditioning is an
issue. Simulations can give an answer in individual cases, and a
minimum time and frequency interval for the solution estimate can be
estimated. However, this time interval can be large, and the true
underlying process can vary significantly within that interval.

\subsection{Tracking versus solving}
\label{sec:FiltVsSolv}


Another important consideration is the statistical method used by the
algorithm to estimate the parameters. Most existing Jones-based and
Physics-based solvers minimize a chi-square.
This is done by using the Gauss-Newton, gradient descent,
or Levenberg-Marquardt algorithm. More recently, in order to solve for
larger systems in the context of calibration of direction dependent
effects, this has been extended using Expectation Maximization, and
SAGE algorithms \citep{Yatawatta08,Kazemi11}. One well-known problem
is that conventional least square minimization and maximum likelihood
solvers lack robustness in the presence of low signal-to-noise ratios
(the estimated minimum chi-square ``jumps'' in between adjacent data
chunks - while this behaviour is non-physical). In most cases, a
filtering of the estimated solutions (Box car, median, etc) or an interpolation might be
necessary \citep[see for example][]{Cotton95}. In practice, situations
of low SNR combined with the need to perform DDE calibration are not
rare (in the case of LOFAR, ionosphere varies on scale of 30 seconds
while not much flux is available in the field).

In this paper we present a new calibration algorithm which structure
is that of a Kalman filter. Our main aim is to address the stability
and ill-conditioning issues discussed above by using a Physics-based
approach, which (i) decreases the number of free
parameters in the model, and (ii) increases the amount of usable
data. The algorithm structure allows to (iii) use additional physical
priors (time/frequency process smoothness for example), while (iv)
keeping the algorithm computationally cheap. Note that we do not do any
quantitative comparison between least-squares solvers and our
approach. Instead, we focus on describing an implementation of a
non-linear Kalman filter for radio interferometry, and we study its
robustness.

While non-linear least-squares solvers
are {\it iterative}, our algorithm uses a non-linear Kalman filter,
which is a {\it recursive} sequence (see
Sec. \ref{sec:KalmanFilters}). Kalman filters are referred in the
literature as {\it Minimum Mean Square Error} estimators, and instead
of fitting the data at best (least-squares solver), they minimize the
error on the estimate, given information on previous states. In other
words, they can be viewed as ``tracking'' the process rather than
solving for it. An estimated process state vector\footnote{The process
  state vector encodes the states of the instrument, ionosphere,
  beams, etc. It is written as $\textbf{x}$ throughout this paper.}
built from previous recursions, together with a covariance matrix
prior are specified. This way, the filter allows to constrain the
expected location of the true process state along the recursion. Even
when the location of the minimum chi-square jumps between data chunks,
the posterior estimate stays compatible with the prior estimate and
with the data (under the measurement and evolutionary models). As more
data goes though the filter, the process state and its covariance are
updated (and the trace of the covariance matrix decreases in general).


An interesting aspect of our approach is that
we use alternative data domains (Sec. \ref{sec:ReprSec}), which amounts to conducing the
calibration in the image domain. We show that this approach provides
higher robustness. 
We discuss the detail of the implementation and algorithmic costs in
Sec. \ref{sec:Implementation}. An important step for the feasibility
of the approach is to re-factor the filtering steps using the Woodbury
matrix identity (Sec. \ref{sec:Woodbury}). We demonstrate the
efficiency of our algorithms in Sec. \ref{sec:Simulations}, based on
simulation of the clock/ionosphere (Sec. \ref{sec:IonClockSim}) and
pointing error (Sec. \ref{sec:PointingWsrt}) problems. An extended
discussion on the differences between our algorithm and other existing
techniques is given in Sec. \ref{sec:discussion}. An overview of the mathematical notation
is given in Tab. \ref{tab:symbols}.

\newcommand{\MyNewVariable}{7cm}
\begin{table}[t]\footnotesize
  \caption{\label{tab:symbols} Overview of the mathematical notations used throughout this paper}
\begin{tabular}{l@{\hspace{0.1cm}}p{\MyNewVariable}}
\midrule
$\textbf{G}_{pt\nu}$\dotfill&
  The product of the $2\times2$ direction-independent Jones matrices for antenna $p$ at time $t$ and frequency $\nu$.\\
$\textbf{D}_{p\vec{s}t\nu}$\dotfill&
  The product of the $2\times2$ direction-dependent Jones matrices in direction $\vec{s}$ for antenna $p$ at time $t$ and frequency $\nu$.\\
$N$\dotfill&The number of parameters in the model.\\
$M$\dotfill&
  The number of visibility-type data points.\\
$M_i$\dotfill&
  The number of image-type data points.\\
$\PVec$\dotfill&
  Process vector of size $N$, containing the values of the parameters to be estimated.\\
$\textbf{y}$\dotfill&
  Data vector of size $M$.\\
$\textbf{P}$\dotfill&The covariance matrix on the estimated process
  vector (size $N\times N$).\\
$\textbf{Q}$\dotfill&The process covariance matrix of size $N\times N$.\\
$\textbf{R}$\dotfill&The data covariance matrix of size $M\times M$.\\
$\textbf{f}$\dotfill&
  The non-linear evolution operator mapping $\mathbb{R}^N\rightarrow \mathbb{R}^{N}$. It is equivalent to a matrix $\textbf{F}$ when $\textbf{f}$ is linear.\\
$\textbf{h}$\dotfill&
  The non-linear measurement operator mapping $\mathbb{R}^N\rightarrow \mathbb{R}^{2M}$. When $\textbf{h}$ is linear, we note it as a matrix $\textbf{H}$. \\
$(.)_{k-1\mid k-1}$\dotfill&
  The {\it a priori} value of $(.)$ at $k-1$ built at the $k-1$ step.\\
$(.)_{k\mid k-1}$\dotfill&
  The prior of $(.)$ predicted at $k$ from the $k-1$ step (after $(.)_{k-1\mid k-1}$ has been evolved through the $\textbf{f}$ evolution operator).\\
$(.)_{k\mid k}$\dotfill&
  The posterior value of $(.)$ estimated at $k$ using the Kalman gain (after the Kalman gain has been applied to $(.)_{k\mid k-1}$ in the data-domain).\\
$\Bchi^i$\dotfill&
  The $i^{th}$ $\sigma$-point vector of size $N$ in the process domain.\\
$\mathY^i$\dotfill&
  The $i^{th}$ $\sigma$-points propagated in the data domain of size $M$.\\
$\textbf{K}$\dotfill&The Kalman gain matrix of size $N\times M$.\\
\bottomrule
\end{tabular}
\end{table}

\subsection{Radio Interferometry Measurement Equation}
\label{sec:RIME}

To model the complex direction-dependent effects (DDE - station beams,
ionosphere, Faraday rotation, etc), we make extensive use of the
Radio Interferometry Measurement Equation (RIME) formalism, which
provides a model of a generic interferometer \citep[for
extensive discussions on the validity and limitations of the measurement
equation see][]{Hamaker96,Oleg11}. Each of the physical
phenomena that transform or convert the electric field before the
correlation is modeled by linear transformations
(2$\times$2 matrices). If $\vec{s}=[\l,\m,\n=\sqrt{1-\l^2-\m^2}]^T$ is a sky
direction, and $\textbf{M}^{H}$ stands for the Hermitian transpose operator of
matrix $\textbf{M}$,
then the $2\times 2$ correlation matrix $\textbf{V}_{(pq)t\nu}$ between antennas
$p$ and $q$ at time $t$ and frequency $\nu$ can
be written as:


\begin{alignat}{3}
\label{eq:ME}
\textbf{V}_{(pq)t\nu}=\textbf{h}(\textbf{x}) &=&& \textbf{G}_{pt\nu}(\textbf{x})\left(\displaystyle\sum\limits_{\vec{s}}
\textbf{V}^{\vec{s}}_{(pq)t\nu}(\textbf{x})k^{\vec{s}}_{(pq)t\nu}
\right)\textbf{G}^H_{qt\nu}(\textbf{x}) \\ 
\textbf{V}^{\vec{s}}_{(pq)t\nu}(\textbf{x})&=&& \textbf{D}_{p\vec{s}t\nu}(\textbf{x})\textbf{X}_{\vec{s}}\ \textbf{D}^H_{q\vec{s}t\nu}(\textbf{x})
\end{alignat}

\noindent where $\textbf{x}$ is a vector containing the parameters of
the given system (ionosphere state, electronics, clocks, etc), $\textbf{D}_{p\vec{s}t\nu}$ is the product of direction-dependent
Jones matrices corresponding to antenna $p$ (e.g., beam, ionosphere
phase screen and Faraday rotation), $\textbf{G}_{pt\nu}$ is the product of
direction-independent Jones matrices for antenna $p$ (like electronic
gain and clock errors), and 
$\textbf{X}_{\vec{s}}$ is
referred as the {\it sky
  term}\footnote{For convenience, in this section and throughout
  the paper, we do not show the sky term $\sqrt{1-l^2-m^2}$ that
  usually divides the sky to account for the projection of the
  celestial sphere onto the plane, as this has no influence on the
  results.} in the direction $\vec{s}$, and is the true
underlying source coherency matrix [[$\X_p\X^*_q$,$\X_p\Y^*_q$],
  [$\Y_p\X^*_q$, $\Y_p\Y^*_q$]]. 
The scalar term $k^{\vec{s}}_{(pq)t\nu}$ describes the
effect of the array geometry and correlator on the observed phase
shift of a coherent plane wave between antennas $p$ and $q$.
We have $k^{\vec{s}}_{(pq)t\nu}=\exp{\left(-2 i\pi
  \phi(\u,\v,\w,\vec{s})\right)}$, with $[\u,\v,\w]^T$ is the baseline
vector between antennas $p$ and $q$ in wavelength units and
$\phi(\u,\v,\w,\vec{s})=\u\l+\v\m+\w\left(\text{n}-1\right)$.


Although the detailed structure of Eq. \ref{eq:ME} is of fundamental importance, throughout
this paper
it is reduced to a
non-linear operator $\textbf{h}:\mathbb{R}^N\mapsto \mathbb{R}^M$,
where $N$ is the number of free parameters and $M$ is the number of
data points. The operator $\textbf{h}$ therefore maps a vector
$\textbf{x}_k$ parameterizing the Jones matrices and/or sky terms appearing the right-hand side of
Eq. \ref{eq:ME} (the states of the beam, the ionosphere, the clocks,
the sky, etc), and maps it to a vector of visibilities $\textbf{y}_k$
such that $\textbf{y}_k=\textbf{h}(\textbf{x}_k)$. In the following, $\textbf{y}_k$ is the set of
visibilities at the time step $k$ for all frequencies, all baselines, and all polarizations. The choice of mapping
the state space to the measurement space for all frequency for a
limited amount of time (time step $k$) is motivated by the fact that regularity is much stronger on the
frequency axis. For example, the Jones matrices associated with ionosphere or clocks, although
they greatly vary in time, have a very stable frequency behaviour at any given time.

\section{Kalman Filter for non-linear systems}
\label{sec:KalmanFilters}


Non-linear least-squares algorithm only
consider the $\chi^2$ value for the given data chunk. As mentioned
above, this is a problem in (i) low SNR and (ii) ill-conditioned
regimes. For example for (i), if one considers a noisy $\chi^2$
valley, the least-square solution will ``jump'' between each time-frequency bin
due to noise - while this behaviour is obviously non-physical. The
effect (ii) will bring instability as a results of the $\chi^2$ valley
having multiple local minima, or a flat minima.
As explained in Sec. \ref{sec:FiltVsSolv}, Kalman filter provide a
number of advantages allowing in principle to significantly improve
robustness, and minimize the impact of ill-conditioning.


In the following, we assume an evolution operator
$\textbf{x}_{k}=\textbf{f}(\textbf{x}_{k-1})+\textbf{w}_k$ describing
the evolution of the physical quantities underlying the RIME, and a
measurement operator
$\textbf{y}_{k}=\textbf{h}(\textbf{x}_k)+\textbf{v}_k$ generating the
set of measurement for a given process vector $\textbf{x}_k$ (examples for
both $\textbf{f}$ and $\textbf{h}$ are given in Sec. \ref{sec:Simulations}). The random
variables $\textbf{v}_k $ and $\textbf{w}_k$ model the noise and are
assumed to follow normal distributions $\textbf{v}_k \sim
\mathcal{N}(0, \textbf{R}_k)$ and $\textbf{w}_k \sim\mathcal{N}(0,
\textbf{Q}_k)$, where $\textbf{R}_k$ and $\textbf{Q}_k$ are the data
and process covariance matrix respectively. In the following, we name
the {\it predicted-process} and {\it data} domains the
codomains of $\textbf{f}$ and $\textbf{h}$ respectively.

\subsection{Kalman Filter}

The traditional Kalman filter \citep{Kalman60} assumes (a) $\textbf{f}$ and $\textbf{h}$ to
be linear operators (written $\textbf{F}$ and $\textbf{H}$ bellow for
$\textbf{f}$ and $\textbf{h}$ respectively). If the process vector $\textbf{x}_{k-1}$ for the time-step $k-1$ has
$\hat{\textbf{x}}_{k-1\mid k-1}$ estimated mean and
$\textbf{P}_{k-1\mid k-1}$ estimated covariance from the data at step
$k-1$, assuming (b) Gaussian noise in the process and data domains,
$\textbf{x}_{k-1\mid k-1}$ is distributed following
$\textbf{x}_{k-1 \mid k-1}\sim\mathcal{N}(\hat{\textbf{x}}_{k-1\mid k-1},
\textbf{P}_{k-1 \mid k-1})$.

Under the conditions (a) and (b), operators $\textbf{F}$ and $\textbf{H}$ yield Gaussian distributions
in the predicted-process and data domains respectively. Given
$\hat{\textbf{x}}_{k-1\mid k-1}$ and $\textbf{P}_{k-1 \mid k-1}$ the
Kalman filter (i) predicts $\hat{\textbf{x}}_{k\mid k-1}$ and
$\textbf{P}_{k \mid k-1}$ through $\textbf{F}$, and (ii) updates those
to $\hat{\textbf{x}}_{k\mid k}$ and
$\textbf{P}_{k \mid k}$ through $\textbf{H}$ given the data
$\textbf{y}_{k}$.

It can be shown that the mean and covariance of $\textbf{x}_{k-1}$ can
be evolved through $\textbf{F}$ giving $\hat{\textbf{x}}_{k\mid k-1}$
and $\textbf{P}_{k\mid k-1}$ as follows:

\begin{alignat}{2}
\label{eq:KF_xf}
\hat{\textbf{x}}_{k\mid k-1} &= \textbf{F}_{k}\hat{\textbf{x}}_{k-1\mid k-1} \\
\label{eq:KF_Pf}
\textbf{P}_{k\mid k-1} &=  \textbf{F}_{k} \textbf{P}_{k-1\mid k-1} \textbf{F}_{k}^{\text{T}} + \textbf{Q}_{k}
\end{alignat}

\noindent Taking into account the data vector $\textbf{y}_k$ at step $k$, the
updated mean and covariance $\hat{\textbf{x}}_{k\mid k}$ and
$\textbf{P}_{k|k}$ of $\textbf{x}_k$ are estimated through the
calculation of the {\it Kalman gain} $\textbf{K}_k$, and are given by:

\begin{alignat}{2}
\label{eq:KF_Sk}
\textbf{S}_k &= \textbf{H}_k \textbf{P}_{k\mid k-1} \textbf{H}_k^\text{T} + \textbf{R}_k \\
\label{eq:KF_Kk}
\textbf{K}_k &= \textbf{P}_{k\mid k-1}\textbf{H}_k^\text{T}\textbf{S}_k^{-1}\\
\tilde{\textbf{y}}_k &= \textbf{y}_k - \textbf{H}_k\hat{\textbf{x}}_{k\mid k-1}\\
\hat{\textbf{x}}_{k\mid k} &= \hat{\textbf{x}}_{k\mid k-1} + \textbf{K}_k\tilde{\textbf{y}}_k\\
\textbf{P}_{k|k} &= (I - \textbf{K}_k \textbf{H}_k) \textbf{P}_{k|k-1}
\end{alignat}

The estimate $\hat{\textbf{x}}_{k\mid k}$ is
optimal in the sense that $\textbf{P}_{k \mid k}$ is minimized. This approach is extremely powerful for linear-systems, but the radio
interferometry Measurement Equation is highly non-linear (the operator
$\textbf{h}$ in Eq. \ref{eq:ME}). This makes the
traditional Kalman filters to be unpractical for radio interferometry
calibration problem.

\subsection{Unscented Kalman Filter}
\label{sec:UKF}

The Kalman filters fails at properly estimating the statistics of
$\textbf{x}$ essentially because $\textbf{f}$ and/or $\textbf{h}$ are
non-linear, and lead to non-Gaussian distributions in the
predicted-process and data domains described above. 

The Unscented
Kalman Filter \citep[UKF,][]{Julier97,Wan00} aims at properly estimating the
mean and covariance in both those domains by
directly applying the non-linear operators $\textbf{f}$ and
$\textbf{h}$ to ``deform'' the initial Gaussian distribution of
$\textbf{x}$.
In practice, instead of selecting a large number of process vectors
built at random as is done
in Monte-Carlo particle filters for example, the Unscented Transform (UT) scheme selects
a much smaller set of $2N+1$
sigma-points in the process domain in a {\it deterministic}
manner.
Each point is characterized by a
location in the process space and a corresponding weight. The set is
built in such a way that its mean and covariance match the statistics
of the random variable $\textbf{x}$. The points are propagated
through the non-linear operators $\textbf{f}$ and $\textbf{h}$ in the
predicted-process and data domains respectively, and the corresponding
mean and covariance are estimated based on their evolved positions and
associated weights. Using Taylor expansions of $\textbf{f}$ and
$\textbf{h}$, it can be shown that the mean and covariance of the
evolved random variable are correct up to the third order of the
expansion \citep{Julier97}. Errors are introduced by higher order terms, but partial
knowledge on those can be introduced using proper weighting schemes. It is
important to note however, that even thought the mean and covariance
can be properly estimated after applying non-linear operators through
the UT, the Kalman filter still assumes Gaussian statistics of all
random variables to estimate the statistics of $\textbf{x}$.

\subsubsection{$\sigma$-points and associated weights}
\label{sec:weights}

Given a multivariate distribution with covariance $\textbf{P}$ of size $N\times N$, the set of $\sigma$-points are generated in the following way:

\begin{equation}
\label{eq:SigmaPoints}
\tilde{\Bchi_i} = \begin{cases}
 \hat{\textbf{x}} & \qquad \text{for} \ i = 0 \\
 \hat{\textbf{x}} + \left [ \sqrt{ \frac{N}{1-w_0} \textbf{P} } \right ]_{i} & \qquad \text{for} \ i = 1,\ldots,N  \\
 \hat{\textbf{x}} - \left [ \sqrt{ \frac{N}{1-w_0} \textbf{P} } \right ]_{i} & \qquad \text{for} \ i = N+1,\ldots,2N
\end{cases}
\end{equation}

\noindent where $N$ is the number of parameters, $\textbf{P}$ is the process covariance matrix, 
$\left [ \textbf{M} \right ]_{i}$ is the $i^{th}$ column of the matrix $\textbf{M}$. 
The real-valued scalar $w_0$ controls the distance of the $\sigma$-points to the origin. As $N$ increases, the radius of the sphere that
contains the $\sigma$-points increases as well.
As shown in \citet{Julier97} for the errors to be minimized on the mean and covariance estimate, the $\sigma$-points should stay in
the neighborhood of the origin. The $\sigma$-point locations are scaled by a parameter $\alpha$ giving:
 
\begin{equation}
\label{eq:SigmaPoints_alpha}
\Bchi_i=(1-\alpha)\tilde{\Bchi_0}+\alpha\tilde{\Bchi_i}
\end{equation}

When estimating the mean of the evolved distribution, 
the weights associated with the $\sigma$-points are

\begin{equation}
w^m_i =  \begin{cases}
        (w_0+\mu-1)/\mu  & \text{for} \ i = 0 \\
        (1-w_0)/(2N\mu)     & \text{otherwise}
        \end{cases}
\end{equation}

\noindent where $\mu$ is a normalizing constant appearing while computing the Taylor expansion of the non-linear operator. 
When computing the covariance of the $\sigma$-points, the weights are given by

\begin{equation}
w^c_i =  \begin{cases}
        w^m_0+\beta+1-\alpha  & \text{for} \ i = 0 \\
        w^m_i     & \text{otherwise}
        \end{cases}
\end{equation}

\noindent where $\beta$ is an extra parameter that can be used to incorporate additional knowledge on the 
fourth-order term of the Taylor expansion of the covariance. 

\subsubsection{Filtering steps}

A set of $\sigma$-points is generated assuming $\textbf{x}_{k-1\mid
  k-1} \sim \mathcal{N}(\hat{\textbf{x}}_{k-1\mid
  k-1},\textbf{P}_{k-1\mid k-1})$, following the scheme outlined
above (Sec. \ref{sec:weights}). The $\sigma$-points are then propagated through the non-linear evolution operator \textbf{f}

\begin{equation}
\Bchi_{k\mid k-1}^{i} = \textbf{f}(\Bchi_{k-1\mid k-1}^{i}) \quad \text{for } i = 0,\dots,2N
\end{equation}

\noindent and the mean and covariance are estimated as follows:

\begin{alignat}{2}
\label{eq:UKF_xf}
\hat{\textbf{x}}_{k\mid k-1} &=&& \sum_{i=0}^{2N} w_{i}^{m} \Bchi_{k\mid k-1}^{i}\\
\nonumber\textbf{P}_{k\mid k-1} &=&& \sum_{i=0}^{2N} w_{i}^{c}\ [\Bchi_{k\mid k-1}^{i} - \hat{\textbf{x}}_{k\mid k-1}] [\Bchi_{k\mid k-1}^{i} - \hat{\textbf{x}}_{k\mid k-1}]^{T}\\
\label{eq:UKF_Pf}
 &&&+\textbf{Q}_k 
\end{alignat}

If $\mu=\alpha^2$ the expressions of $\hat{\textbf{x}}_{k\mid k-1}$ and $\textbf{x}_{k\mid k-1}$ agree up to the third order
of the Taylor expansion. Note that Eq. \ref{eq:UKF_xf} and \ref{eq:UKF_Pf} are the UKF versions
of Eq. \ref{eq:KF_xf} and \ref{eq:KF_Pf}.

Once $\hat{\textbf{x}}_{k\mid k-1}$ and $\textbf{P}_{k\mid k-1}$ are estimated, we assume 
$\textbf{x}_{k\mid k-1} \sim \mathcal{N}(\hat{\textbf{x}}_{k\mid k-1},\textbf{P}_{k\mid k-1})$, and a new set of 
$\sigma$-points is generated following the scheme outlined in Eqs. \ref{eq:SigmaPoints} and \ref{eq:SigmaPoints_alpha}.

The new set of $\sigma$-points are propagated onto the measurement domain through the non-linear observation function \textbf{h}

\begin{equation}
\label{eq:PredictStepUKF}
\mathY_{k}^{i} = \textbf{h}(\Bchi_{k\mid k-1}^{i}) \quad \text{for } i = 0,\dots,2N
\end{equation}

\noindent where $\mathY_{k}^{i}$ is the measurement vector corresponding to each process vector $\Bchi_{k\mid k-1}^{i}$. 
As in Eq. \ref{eq:UKF_xf} and \ref{eq:UKF_Pf} the measurement mean
$\hat{\textbf{y}}_{k}$, measurement covariance $\textbf{P}_{y_{k}y_{k}}$, and state-measurement cross-covariance $\textbf{P}_{x_{k}y_{k}}$
are then estimated:

\begin{alignat}{2}
\label{eq:UKF_xh}
\hat{\textbf{y}}_{k} &=&& \sum_{i=0}^{2N} w_{m}^{i} \mathY_{k}^{i} \\
\label{eq:UKF_Pzzh}
\textbf{P}_{y_{k}y_{k}} &=&& \sum_{i=0}^{2N} w_{c}^{i}\ [\mathY_{k}^{i} - \hat{\textbf{y}}_{k}] [\mathY_{k}^{i} - \hat{\textbf{y}}_{k}]^{T} + \textbf{R}_k\\
\label{eq:UKF_Pxyh}
\textbf{P}_{x_{k}y_{k}} &=&& \sum_{i=0}^{2N} w_{c}^{i}\ [\Bchi_{k\mid k-1}^{i} - \hat{\textbf{x}}_{k\mid k-1}] [\mathY_{k}^{i} - \hat{\textbf{y}}_{k}]^{T}
\end{alignat}

\noindent Note again that Eq. \ref{eq:UKF_xh} and \ref{eq:UKF_Pzzh}
mirror the behaviour of Eq. \ref{eq:KF_Sk}, while the term
$\textbf{P}_{x_{k}y_{k}}$ (Eq. \ref{eq:UKF_Pxyh})
is similar to $\textbf{P}_{k\mid k-1}\textbf{H}_k^\text{T}$ of Eq. \ref{eq:KF_Kk}. $\hat{\textbf{y}}_{k}$ has size $M$,
$\textbf{P}_{y_{k}y_{k}}$ has size $M\times M$ and $\textbf{P}_{x_{k}y_{k}}$ has size $N\times M$, where $N$ and $M$ are 
the dimensions of the process and measurement spaces respectively. The Kalman gain is then

\begin{equation}
\label{eq:UKF_K}
\textbf{K}_{k} = \textbf{P}_{x_{k}y_{k}} \textbf{P}_{y_{k}y_{k}}^{-1}
\end{equation}

\noindent and the updated estimates $\hat{\textbf{x}}_{k\mid k}$ and $\textbf{P}_{k\mid k}$ can be computed:

\begin{alignat}{2}
\label{eq:UKF_final}
\hat{\textbf{x}}_{k\mid k} &=& \hat{\textbf{x}}_{k\mid k-1} + \textbf{K}_{k}( \textbf{y}_{k} - \hat{\textbf{y}}_{k} )\\
\textbf{P}_{k\mid k} &=& \textbf{P}_{k\mid k-1} - \textbf{K}_{k} \textbf{P}_{y_{k}y_{k}} \textbf{K}_{k}^{T}
\end{alignat}

\def\mathR{\bm{\mathcal{R}}}

\begin{figure*}[]
\begin{center}
\includegraphics[width=16cm]{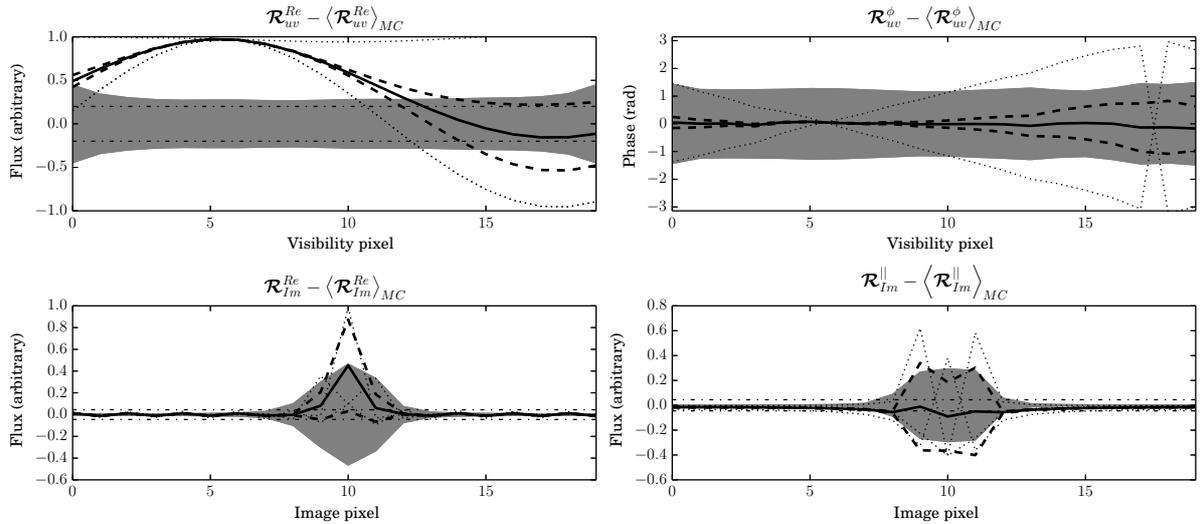}
\caption{\label{fig:ReprSig} We use Monte-Carlo simulations to study
  the ability of the Unscented Transform to properly describe the data
  statistics after the non-linear measurement operator and the
  different representation operators $\mathR$
  (Sec. \ref{sec:ReprSecIntro}) have been applied to the process
  domain. Here we consider the case of a frequency-dependent phase
  gradient (clock offset, ionospheric effect, or source position
  offset). We have subtracted the true mean $\left<\mathR\right>_{MC}$ to each quantity plotted
  in this figure. The true covariance corresponds to the gray area,
  while the sigma points appear as dotted lines, together with their
  associated estimated covariance (thick dashed line). The noise in
  the data is represented by the dash-dotted line. Qualitatively, the
  goodness of the description depends on the type of the data
  representations operator.}
\end{center}
\end{figure*}

\section{Data representation}
\label{sec:ReprSec}

In this section we describe how we can modify the measurement operator
together with the raw data to improve robustness. Using the operator
discussed in Sec. \ref{sec:1Dimage} in combination with the Kalman
filter discussed above, this is equivalent to an image-plane calibration.

\subsection{Robustness with large process covariance}
\label{sec:ReprSecIntro}

As explained in Sec. \ref{sec:UKF} and
\ref{sec:Implementation}, the Unscented Transform correctly
approximates the evolved covariance up to the third order. When the
radius of the sphere containing $\sigma$-points in the process domain
increase, and depending on the strength of non-linearities of the
evolution and measurement operators \textbf{f} and \textbf{h}, the
estimated mean and covariance can be affected by large errors. This is
the case when the multivariate ellipsoid is too deformed, and the
statistics of the $\sigma$-points in the evolved domain do not capture
anymore the true statistics.

Here, we introduce another layer of non-linear operators
$\mathR:\mathbb{R}^{2M}\mapsto \mathbb{R}^{\tilde{M}}$ that transform the sets of visibilities into another
measurement domain. 
We define various simple representation operators as follows:

\def\RepUV{\bm{\mathcal{R}}^{\mathbb{C}}_{\mathrm{uv}}}
\def\RepUVRe{\bm{\mathcal{R}}^{Re}_{\mathrm{uv}}}
\def\RepUVPhi{\bm{\mathcal{R}}^{\phi}_{\mathrm{uv}}}
\def\RepIm{\bm{\mathcal{R}}^{\mathbb{C}}_{\mathrm{Img}}}
\def\RepImRe{\bm{\mathcal{R}}^{Re}_{\mathrm{Img}}}
\def\RepImAbs{\bm{\mathcal{R}}^{||}_{\mathrm{Img}}}
\def\RepaRe{\bm{\mathcal{R}}^{\mathbb{C}}_{\mathrm{Img}}}
\def\RepbAbs{\bm{\mathcal{R}}^{||}_{\mathrm{Img}}}
\def\OpIm1D{\textbf{g}_{\text{1D}}}
\def\VisVec{\textbf{y}}

\begin{alignat}{5}
\nonumber\RepImAbs:& \mathbb{R}^{2M}\rightarrow \mathbb{R}^{M_i} 
&\qquad
          \RepUVRe:& \mathbb{R}^{2M}\rightarrow \mathbb{R}^M\\
\nonumber          & \VisVec\mapsto \left|\OpIm1D(\VisVec)\right|
&\qquad
                   & \VisVec\mapsto Re(\VisVec)\\
\nonumber \RepUVPhi  :& \mathbb{R}^{2M}\rightarrow \mathbb{R}^{M} &
\qquad
\RepImRe:& \mathbb{R}^{2M}\rightarrow \mathbb{R}^{M_i}\\
\nonumber          & \VisVec\mapsto \phi(\VisVec) &
\qquad
         & \VisVec\mapsto Re(\OpIm1D(\VisVec))
\end{alignat}

\noindent where $\OpIm1D=\RepIm$ is the operator described in
Sec. \ref{sec:1Dimage} transforming a set of visibilities into another
set of 1-dimensional images (the Fourier transform along the frequency
axis, also refered later as the
{\it pseudo-image} domain). The operators $\left|\textbf{.}\right|$, $\phi(\textbf{.})$ and
$Re(\textbf{.})$ take the complex norm, the phase and the real
part respectively. $M$ is the number of visibility
data, $M_i$ the number of pixels in the image domain. 
Our goal is to obtain a system that has less
non-linearity, so that the $\sigma$-points statistics still properly
match the true statistics, even when the volume within the multivariate ellipsoid is
large in the process domain.



In order to illustrate this idea, we compare the evolved covariance
estimated using Eq. \ref{eq:UKF_Pzzh} to the true evolved mean and
covariance as estimated by running Monte Carlo simulations. The system
is made of one point source at the phase center, the bandpass goes
from $30$ to $70$ MHz, and our interferometer consists of one
baseline. We consider the ellipsoid of the clock-offset parameter
$\Delta t_{01}\sim \mathcal{N}(0,10)\times 10^{-9}\text{s}$
(therefore corresponding to a large $10\times 10^{-9}$ sec. estimated
error), and inspect how it is reflected in the data domains after
applying $\widetilde{\textbf{h}}=\mathR\circ\textbf{h}$ (the symbol
$\circ$ is used here for the function composition). For that system,
Eq. \ref{eq:ME} becomes
$\widetilde{\textbf{y}}=\widetilde{\textbf{h}}(\textbf{x})=\mathR\left(\exp{\left(2\pi
  i \nu \Delta t_{01}\right)}\right)$, where $\Delta t_{01}$ is our
random variable.
Fig. \ref{fig:ReprSig} shows how the statistics of the $\sigma$-points
compare to the true statistics when using different data
representations $\mathR$. When $\mathR$ picks the real part $Re$ or
the phase $\phi$ of $\textbf{y}=\widetilde{\textbf{h}}(\textbf{x})$
($\RepUVRe$ and $\RepUVPhi$), the $\sigma$-points statistics are
obviously wrong. 

\subsection{One-dimensional image domain for calibration}
\label{sec:1Dimage}

Although a single transformation separate the uv-plane from the image
domain, it seems the later is sometimes more suited for
calibration. Intuitively, in the uv domain, clock shifts, source
position, ionospheric disturbance will wrap the phases of the
complex-valued visibilities {\it
  everywhere}, and the strong non-linearities sometimes present in
\textbf{h} make the distribution strongly non-Gaussian. On the
contrary, in the image domain, the same type of perturbations only
affect the data {\it locally}, and will move the flux from one pixel
to its neighborhood.



We cannot use the 2-dimensional image domain, as it is build from the
superposition of all baselines, which would lead to
an effective loss of information. Instead, similarly to what is done
for VLBI delays and fringe rates calibration \citep{Cotton95}, we build a 1D
image per baseline (see Appendix \ref{sec:1Ddetail} for details). As shown in Fig. \ref{fig:ReprSig}, when going into the pseudo-image domain, the power is
concentrated in a few pixels. The real part $\RepImRe$ of individual
pixels gives a better match, but is still biased. Taking the norm
$\RepImAbs$ of the image-plane complex pixel seems to behave well in
all conditions. This is discussed in detail in Appendix \ref{sec:1Ddetail}.


\subsection{The augmented measurement state for regularization}
\label{sec:ImplAMM}


As explained above, one of the aim of the work presented in this paper
is to address the ill-conditioning issues related to the large inverse
problem underlying the use of modern interferometers. This is done by
analytically specifying the physics underlying the RIME (using a
Physics-based approach - see Sec. \ref{sec:PhysPeal}), and by using
the Kalman filter mechanism, able to constrain the location of the
true process state through the transmission of previous estimated state
and associated covariance. Yet, in some situations, and particularly when two
variables are analytically degenerate (such as the clock shifts and
the ionosphere when the fractional bandwidth is small), the robustness of the scheme
presented is not strong enough to
guarantee the regularity of solutions, and the estimated process can
drift to a domain where solutions are non-physical.



In order to take into account external constrains while still properly evolving
the process covariance $\textbf{P}_{k\mid k}$, we introduce an
augmented measurement model \citep[see for example][]{Henar09,Hiltunen10}. 
Using the idea underlying Tikhonov regularization, if $\textbf{x}_0$
is the expected value of $\textbf{x}$, and $\textbf{Q}^\gamma_0$ the covariance of $\textbf{x}_0$, our cost function
becomes:

\begin{alignat}{3}
\mathcal{C}(\textbf{x})&=\| \widetilde{\textbf{h}}(\textbf{x})-\widetilde{\textbf{y}} \|^2_\textbf{P} + \|\textbf{x}-\textbf{x}_0 \|^2_{\textbf{Q}^\gamma_0}\\
&=
\left \|
\left [
\begin {array}{c}
\widetilde{\textbf{h}}(\textbf{x})\\ 
\textbf{x}
\end {array}
\right ]
-
\left [
\begin {array}{c}
\widetilde{\textbf{y}}\\ 
\textbf{x}_0
\end {array}
\right ]
\right \|^2_{(\textbf{P},\textbf{Q}^\gamma_0)} \\
&=\| \widetilde{\textbf{h}}^a(\textbf{x})-\widetilde{\textbf{y}}^a \|^2_{(\textbf{P},\textbf{Q}^\gamma_0)}
\end{alignat}

\noindent where
$\|\textbf{x}\|_{\textbf{C}}=\textbf{x}^T\textbf{C}^{-1}\textbf{x}$,
is the norm of vector \textbf{x} for the metric
\textbf{C}, with \textbf{C} the covariance matrix of \textbf{x} (or
the Mahalanobis distance). The operator
$\widetilde{\textbf{h}}^a:\mathbb{R}^N\mapsto \mathbb{R}^{M+N}$ is the augmented
version of $\textbf{h}$ and $(\textbf{P},\textbf{Q}^\gamma_0)$ is the block diagonal
covariance matrix of the augmented process vector. The parameter $\gamma$ allows to
control the strength of the Tikhonov regularization, and is such that
$\textbf{Q}^\gamma_0=\gamma^{-2}\textbf{Q}_0$.



\section{Implementation for radio-interferometry}
\label{sec:Implementation}

As explained above, radio-interferometry deals with a large inverse
problems, made of millions or billions of non-linear equations. This
poses a few deep problems including (i) numerical cost and (ii) numerical stability. In
this section, we describe our UKF implementation.


\subsection{Woodbury matrix identity}
\label{sec:Woodbury}

The first issue is the size of the matrices involved in the UKF
recursion steps presented in Sec. \ref{sec:UKF}. Specifically, in the
case of LOFAR, we have $n_{bl}\sim 1500$ baselines and $n_\nu\sim 250$
frequencies. This gives a number of
dimensions $M$ for the measurement space of $M\sim1.5 \times10^6$ per recursion
(taking into account the 4-polarization visibilities). The predicted
measurement covariance matrix $\textbf{P}_{y_{k}y_{k}}$ has size
$M\times M$, and in practice becomes impossible to store and invert directly ($\sim8$ Peta-bytes of 
memory).
Fortunately we can re-factor Eq. \ref{eq:UKF_K} so that we do not have to
explicitly calculate each cell of $\textbf{P}_{y_{k}y_{k}}$. We can see that Eq. \ref{eq:UKF_Pzzh} can be rewritten as


\begin{alignat}{3}
\label{eq:UKF_Pzzh_S}
\textbf{P}_{y_{k}y_{k}} &=&& \ \textbf{S}_k\textbf{W}\textbf{S}_k^T + \textbf{R}_k\\
\text{with } &&&
\begin{cases}
[\textbf{S}_k]_i      &= \ \mathY_{k}^{i} - \hat{\textbf{y}}_{k}\\
\textbf{W}_{ij}  &= 
\begin{cases}
  w^c_i & \qquad \text{if} \ i = j \\
  0 & \qquad \text{otherwise} \\
\end{cases}
\end{cases}\nonumber
\end{alignat}

\noindent where $\textbf{S}_k$ is a matrix of size $M\times (2N+1)$,
$[\textbf{S}_k]_i$ is the $i^{th}$ column of $\textbf{S}_k$, and $\textbf{W}$
is a diagonal matrix of size $(2N+1)\times (2N+1)$ containing the weights on its
diagonal. Using the Woodbury matrix
identity\footnote{\def\A{\textbf{A}}
\def\U{\textbf{U}}
\def\V{\textbf{V}}
\def\C{\textbf{C}}
The Woodbury Matrix Identity has sometimes been used in the context of the
Ensemble Kalman Filters, and is given by:

$\left(\A+\U\C\V \right)^{-1} = \A^{-1} - \A^{-1}\U \left(\C^{-1}+\V\A^{-1}\U \right)^{-1} \V\A^{-1}$
} \citep{Hager89}, we can express the Kalman gain $\textbf{K}_{k}$:

\begin{alignat}{2}
\label{eq:Pzz_inv_Woodbury}
\nonumber\textbf{K}_{k} &= \textbf{P}_{x_{k}y_{k}}\textbf{P}_{y_{k}y_{k}}^{-1} \\
&=\textbf{P}_{x_{k}y_{k}}
\MR^{-1}\left(\textbf{I} - \MS \left( \MW^{-1}+\MS^T \MR^{-1} \MS \right) \MS^T \MR^{-1}\right)
\end{alignat}

This relation (Eq. \ref{eq:Pzz_inv_Woodbury}) is quite remarkable, as
it allows us to apply the Kalman gain without explicitly calculating
it, and without estimating $\textbf{P}_{y_{k}y_{k}}$ and its inverse
either. Instead, the inverse $\MW^{-1}$ of the diagonal weight matrix of size
$(2N+1)\times (2N+1)$, and the inverse $\MR^{-1}$ of the data
covariance matrix of size $M\times M$ have to be estimated. Even
though $\MR$ has large dimensions, if the noise is uncorrelated only
the diagonal has to be stored and the inverse can be computed
element-by-element. Similarly, inner product of matrices with $\MR$ are
computationally cheap. At each recursion $k$ we have to explicitly estimate
the $\sigma$-points evolved through the measurement
operator \textbf{h} and contained in $\textbf{S}_k$.

\subsection{Adaptive step}

While $\textbf{P}_{k\mid k}$ characterizes the posterior
process covariance, the matrix $\textbf{Q}$ (Eq. \ref{eq:KF_Pf} and \ref{eq:UKF_Pf}) characterizes the
{\it intrinsic} process covariance through time. It can for example describe the natural
time-variability of the ionosphere, the speed of the clock drift, or
the beam stability. 
In addition, in strong non-linear regime, it is well known that the Kalman filters can
underestimate $\textbf{P}_{k\mid k}$, and thereby drive biases in the
estimate $\hat{\textbf{x}}_{k\mid k}$ of $\textbf{x}_k$. This would
typically happen when $\hat{\textbf{x}}_{k-1\mid k-1}$ is too far from
$\textbf{x}$, or when the process covariance is changing
(for example a changing and increasing ionospheric disturbance for a given time-period).
Although the Kalman filter does not produce an update $\textbf{Q}_k$ of
$\textbf{Q}$, based on the residual data we can externally update it
and write $\textbf{Q}_k=\kappa\textbf{Q}$. The scaling factor $\kappa$ is
useful to estimate whether the model is properly fitting the data at
any time step $k$. Following \citet{Ding07}, we can write:



\begin{alignat}{3}
\kappa&=&&
\sum\limits_{i}w^i_{\textbf{Q}}\frac{
\operatorname{tr}\left\{
(\textbf{y}_{k-i}-\hat{\textbf{y}}_{k-i})(\textbf{y}_{k-i}-\hat{\textbf{y}}_{k-i})^T-\textbf{R}_{k-i}
\right\}
}{
\operatorname{tr}\left\{
\textbf{S}_{k-i}\textbf{W}\textbf{S}_{k-i}^T
\right\}
}\\
w^i_{\textbf{Q}}&=&&\exp{\left(-(t_k-t_{k-i})/\tau_{\textbf{Q}_k}\right)} 
\end{alignat}

\noindent where $\operatorname{tr}\left\{\textbf{A}\right\}$ is the operator computing the trace of a
matrix $\textbf{A}$. The weights are designed to take into account past residual
values, and $\tau_{\textbf{Q}_k}$ is a time-type
constant. Here, estimating $\kappa$ is computationally cheap as $\operatorname{tr}$
only accesses the diagonal of the input matrix.

\begin{figure*}[t!]
\begin{center}
\includegraphics[width=17cm]{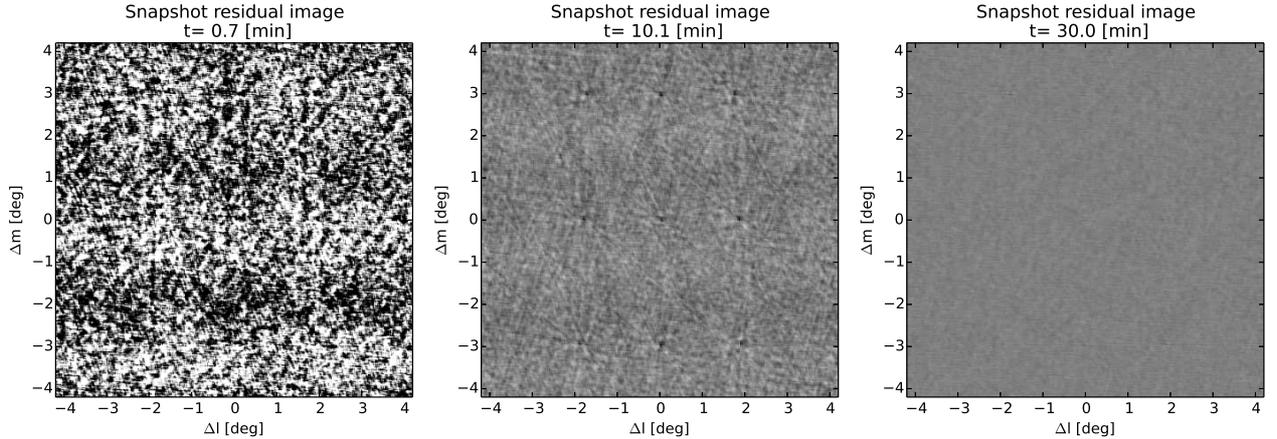}
\caption{\label{fig:ImResid}
  This figure shows the snapshot residual image estimated at different
  recursion times (the color scale is identical in each panel). As
  recursion time evolves, more data have ``crossed'' the Kalman filter,
  the ionospheric and clock parameters estimates are getting more
  accurate, and the residual noise level decreases (see also Fig. \ref{fig:SimsResid}).
}
\end{center}
\end{figure*}

\subsection{Computational cost}


In this section, we discuss the computational cost of the proposed
algorithm. Our concern is to show that the approach is feasible, and we do not intend to show that it is faster
than other existing approaches.
However, we discuss the issues of the scaling relations and parallelizability of
various parts of the algorithm. We argue that using the
refactorization described in Sec. \ref{sec:Woodbury}, our algorithm
should be compatible with existing hardware, even for the datasets
produced by the most modern radiotelescopes.


As explained above, 
we adopt a Physics-based approach to reduce the number of degrees of
freedom by orders of magnitudes while using more data at a time (see
Sec. \ref{sec:PhysPeal} for a discussion on Jones-based versus
Physics-based approach). The Kalman filter is fundamentally recursive
(i.e. it has only one iteration), while tens to hundreds of iterations
are needed to reach local convergence with Levenberg-Marquardt for
example. This means that the equivalent gain by the proposed approach
on the model estimation side is the
number of iteration. This gain might be balanced in some cases by the
larger data chunks processed at a time by the Kalman filter itself. We
give here the scaling relations for our implementation of the filter
scheme.

The predict step (controlled by operator $\textbf{f}$ in Sec. \ref{sec:UKF}) always
represents a relatively small number of steps, as it scales with the
number of parameters $N$ in the process vector.
The update step however is the costly part of the computation
(Sec. \ref{sec:UKF}). It consists of (i) estimating the data
corresponding to different points in the process domain (applying the
operator $\textbf{h}$, see Eq. \ref{eq:ME}) and (ii) computing the
updated estimate of the process vector and associated covariance.
Step (i) is common to {\it all} calibration algorithms, and in the
majority of cases this is the expensive part of the calculation as we
map $N$ parameters to $M$ data points, $M$ having relatively high
values of $\sim10^4-10^6$. Indeed, within our framework, most of the
computation is spent in the estimate of $\textbf{S}_k$ of size $M\times
(2N+1)$ (Eq. \ref{eq:PredictStepUKF}), which, compared to the Jacobian
equivalent that would have size $M\times N$, represents a cost
of a factor of $\sim2$ in computational cost. It is worth to note that this step is heavily
parallelisable.
The series of operation in (ii) to apply the Kalman gain
(Eq. \ref{eq:Pzz_inv_Woodbury}) are negligible in terms of
computing time for the example described in
Sec. \ref{sec:IonClockSim}. From the scaling associated with the use
of the Woodbury matrix identity, following the work presented in
\citet{Mandel06}, we estimate this operation should scale as
$\mathcal{O}(N^3 + MN^2)$.


We believe the algorithm should be practical with large datasets. For
the few test cases we have been working on (with a 4-core CPU) with
$\sim20$ to $\sim100$ parameters, and moderate to large dataset (such
as the LOFAR HBA dataset containing $\sim1500$ baselines, 30
sub-bands, and a corresponding $M\sim1.8\times 10^5$ points {\it per
  recursion}, see Sec. \ref{sec:SimulLOFAR}), the algorithm was always
faster than real time by a factor of several. On the LOFAR CEP1
cluster, a distributed software would work with up to $M\sim 10^6$ points per recursion and $N\sim100$. Preliminary tests showed
that the Kalman filter most consuming steps appearing in
Eq. \ref{eq:Pzz_inv_Woodbury} are computed within a few seconds
(computing inner products with numpy/ATLAS, on an 8-core CPU).


\section{Simulations}
\label{sec:Simulations}


In this section, we present simulations for (i) the pointing
error calibration problem \citep[also addressed using Physics-based
  algorithms in][]{SB:pointing,Smirnov2011qmc} and (ii) the clock/ionosphere problem. 


\subsection{Clock drifts and ionosphere}
\label{sec:IonClockSim}


LOFAR raw datasets are characterized by a few dominating
direction-independent and direction-dependent effects, including
clocks and ionosphere. While direction dependent calibration is known
to be difficult, clock errors and ionosphere effects combined with a
limited - even though large - bandwidth make the problem partially
ill-conditioned.

\subsubsection{Evolution and Measurement operators}
\label{sec:SimulOper}

The error $\delta t^{\mathrm{clk}}_p$ due to the clock offset of antenna $p$
produce a linearly frequency-dependent phase $\phi^{\mathrm{clk}}_p=2\pi \nu \delta
t^{\mathrm{clk}}_p$. The time delay $\delta t_{p,d}^{\mathrm{ion}}$ introduced by the
ionosphere is frequency dependent $\delta t_{p,d}^{\mathrm{ion}}\propto
T_{p,d}/\nu^2$, where $T_{p,d}$ is the Total Electron Content (TEC),
given in TEC-units (TECU), and seen by station $p$ in direction
$d$. This gives a phase $\phi_{p,d}^{\mathrm{ion}}=k T_{p,d}/\nu$, with
$k=8.44\times 10^9$ m$^3$.s$^{-2}$.



\def\TECVec{\textbf{T}^\textbf{d}_{\textbf{p}}}
\def\ClockVec{\bm{\delta}\textbf{t}_{\textbf{p}}}
\def\TECVec{T^{d}_{p}}
\def\ClockVec{\delta t_p}

The measurement operator $\textbf{h}$ (Eq. \ref{eq:ME}) is built
from the direction independent $\textbf{G}_{pt\nu}$ and direction dependent
terms $\textbf{D}^d_{pt\nu}$ as follows:

\begin{alignat}{3}
\textbf{G}_{pt\nu}(\PVec)&:=&&\exp\left(2\pi i \nu\ \ClockVec(\PVec)\right)\textbf{I}\\
\textbf{D}^d_{pt\nu}(\PVec)&:=&&\exp\left(i k\nu^{-1}\ \TECVec(\PVec) \right)\textbf{I}
\end{alignat}

\noindent where $\textbf{I}$ is the $2\times 2$ unity matrix,
$\ClockVec$ is a simple linear operator unpacking the clock value of
antenna $p$ from the process vector $\PVec$. For this test, we choose
to model the ionosphere using a Legendre polynomial function
\citep[see for example][]{Yavuz07}. Assuming a single ionospheric
screen at a height of $100$ km, the non-linear operator $\TECVec$
extracts the Legendre coefficients, and returns the TEC value seen by
antenna $p$ in direction $d$. For this simulation, we are using the
$\RepImAbs$ representation presented in Sec. \ref{sec:ReprSec}.

The operator $\textbf{f}$ describing the dynamics of the system
typically contains a lot of physics. Clock
offsets drift linearly with time, while the ionosphere has a
non-trivial behaviour (defined in Sec. \ref{sec:SimulLOFAR}). 
We configure the filter to consecutively use two types of evolution
operator $\textbf{f}$. The first is used for the first $\sim6$
minutes, and $\textbf{f}$ is the identity function, which corresponds
to a stochastic evolution. This appears useful when the
initial process estimate starts far from the true process state. This
way, the filter's state estimate get closer to the true state without
assuming any physical evolutionary dynamics. The convergence speed and
accuracy are then both controlled by the covariance matrices
$\textbf{P}_{k}$ and $\textbf{Q}_k$ described above.  We set the
second evolutionary operator $\textbf{f}$ to be an extrapolating
operator. It computes an estimated process vector value from the past estimates. At time step $k$, for the $i^{th}$ component of $\PVec$,
solutions are estimated as follows:

\begin{alignat}{3}
\label{eq:EvolOp}
\textbf{f}(\PVec^i_k)&:=&\bm{\mathcal{P}}\left(
\{\PVec_{k-m}\}, n^i_{\mathrm{fit}}, \tau_i
\right)\\
w^i_{t-m}&=&\exp{\left(-(t_k-t_{t-m})/\tau_i\right)} 
\end{alignat}

\noindent where $\bm{\mathcal{P}}$ is the operator computing the
polynomial interpolation, $n^i_{\mathrm{fit}}$ is the degree of the
polynomial used for the interpolation,
$w_i$ are the weights associated with each point $\PVec_{k-m}$ at
$t_{k-m}$, and $\tau_i$ gives a time-scale beyond which past
information is tapered. 

\subsubsection{Simulation for LOFAR}
\label{sec:SimulLOFAR}

An important consideration is that at any given time our algorithm needs to
access all frequencies simultaneously. With $250$ sub-bands ($16$-bit
mode), $1$ channel per sub-band, $1500$ baselines, $4$ polarization,
this gives a number of visibilities per recursion of $1.5\times 10^6$. The data is currently distributed and stored per
sub-band - so our software needs to deal with a number of technical
issues for a realistic full size simulation. For this
simulation, we scale down the problem by a factor $\sim 8$ in terms of
number of frequency points per recursion. Assuming NVSS source counts \citep{Condon98}, a spectral index
of $-0.8$, and a field of view of $8$ degrees in diameter, we estimate
a total of $\sim 20$ Jy of signal per pointing at $\sim 150$
MHz. Inspecting the cross polarizations visibilities of a LOFAR
calibrated data-set with $\Delta \nu=0.2$ MHz and $\Delta t=6$ s gives
an estimated noise of $\sim 2$ Jy per visibility bin. We work on $30$ sub-bands
only, with frequencies linearly distributed between $100$ and
$150$ MHz, and scale the signal to match $\mathrm{SNR}\sim 10$ per visibility
per sub-band. We distribute the corresponding flux density on a
$3\times 3$ rectangular grid of sources with a step of $3$ degrees in
RA and $3$ degrees in DEC.

For the dynamics of the underlying physical effects, we apply a
linearly drifting clock offset taken at random with $\partial(\delta
t^{\mathrm{clk}}_p)/\partial t \sim\mathcal{N}(0, 10)$ ns. As mentioned above
we model the ionosphere with a 2D-Legendre polynomial basis function. For this
simulation, each Legendre $l_{ij}$ coefficients varies following
$l_{ij}=\sin (t/\tau_{ij}+d\tau_{ij})$, with $\tau_{ij}$ and
$d\tau_{ij}$ taken at random. Along these lines discussed above, we
set $(n^i_{\mathrm{fit}},\ \tau_i)=(1,\ 5 \text{min})$ and
$(n^i_{\mathrm{fit}},\ \tau_i)=(2,\ 1 \text{min})$ for the clock
and for the ionosphere respectively. These orders for the extrapolating
polynomials are in agreement with the linear clock drift, and the
non-linear behaviour of the ionosphere.

\begin{figure*}[]
\begin{center}
\includegraphics[width=17cm]{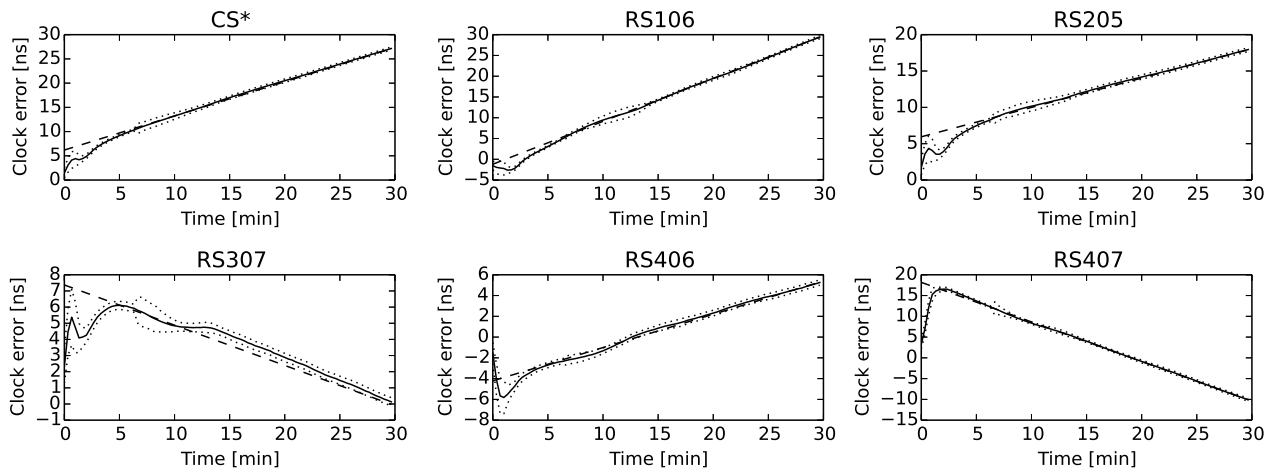}
\caption{\label{fig:SimsClocks} 
This figure shows the estimated clock
  errors (full line) as well as the posterior covariance (dotted line)
  as a function of time for different LOFAR stations in the simulation presented in
  Sec. \ref{sec:IonClockSim}. The dashed line shows the true
  underlying clock offset. In order to improve convergence speed, before $t=6$ min. the evolutionary model
  is stochastic.}
\end{center}
\end{figure*}

\subsubsection{Results}

\begin{figure}[]
\begin{center}
\includegraphics[width=8.5cm]{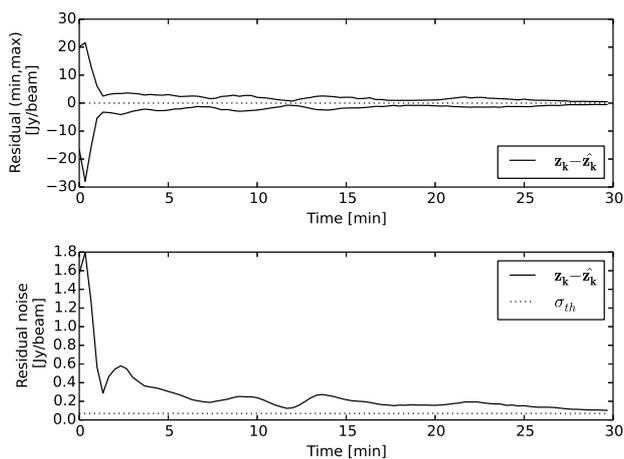}
\caption{\label{fig:SimsResid} Top panel shows the maximum and minimum
residual values in the snapshot image as a function of recursion time (full line). In the bottom
panel we plot the standard deviation in the residual snapshot
maps. The expected thermal noise is shown in the bottom figure as the
dotted line.}
\end{center}
\end{figure}

\begin{figure}[]
\begin{center}
\includegraphics[width=8.5cm]{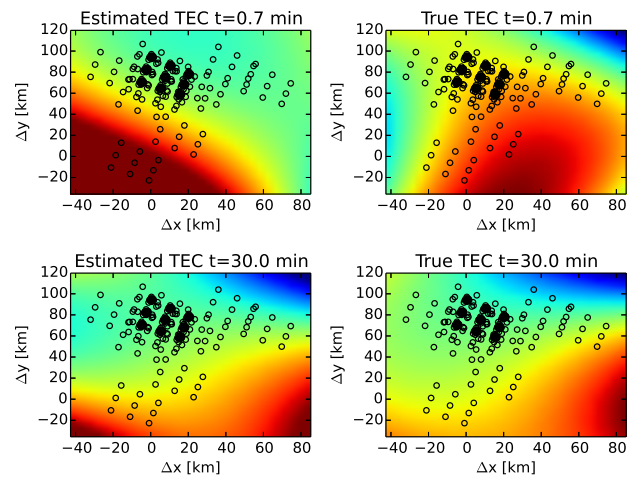}
\caption{\label{fig:SimsIon} The ionosphere state is described using a
Legendre polynomial basis function. We show here the estimated and
true state of the ionosphere (left and right panels), at the beginning
and at the end of the filter's recursion. The open circles show the
location of the pierce points in the ionosphere. The spacial coordinates are
given in kilometers from the array center projected on the ionosphere screen.}
\end{center}
\end{figure}

The filter and simulation configuration are described in
Sec. \ref{sec:SimulOper} and \ref{sec:SimulLOFAR} respectively. At
each recursion step, $\sim 180.000$ complex visibilities ``cross'' the filter and the
process state (clock and ionosphere) as well as its covariance
are estimated.

First, in order to inspect the match to the data we derive the
frequency-baseline-direction dependent Jones matrices at the discrete
locations of the sources in our sky-model, and subtract the
visibilities corresponding to the given discrete directions. We then
grid the residual visibilities and compute the snapshot
images (see Fig. \ref{fig:ImResid}). Fig. \ref{fig:SimsResid} shows the minimum and maximum
residual as well as the standard deviation as a function of time. Very
quickly the visibilities are correctly matched down to the thermal
noise.

\begin{figure*}[ht!]
\begin{center}
\includegraphics[width=17cm]{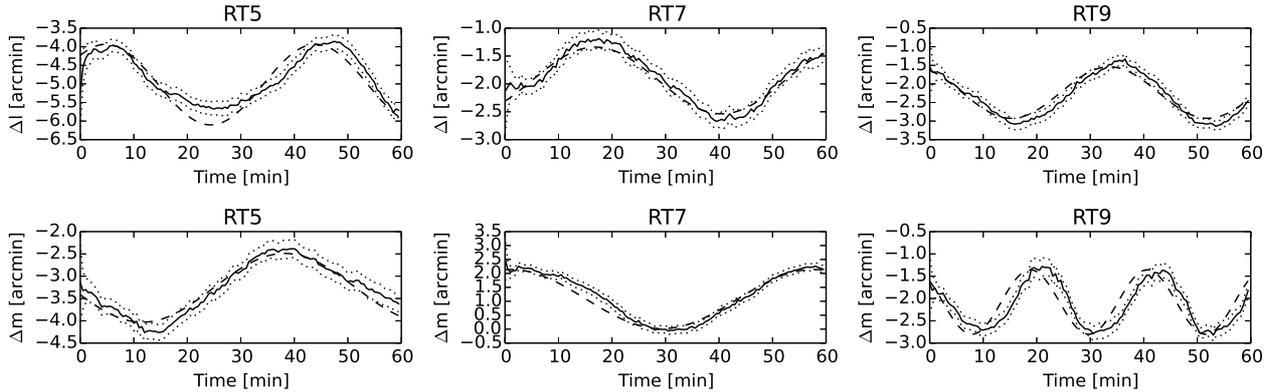}
\caption{\label{fig:SimsWsrt} This figure shows the true pointing
  errors for a WSRT simulation (dashed line) together with the
  estimated state as a function of time (full line). Our algorithm
  properly tracks the time dependent pointing errors within the
  estimated covariance (dotted line).
  }
\end{center}
\end{figure*}

In Fig. \ref{fig:SimsClocks} we show the clock offsets estimates as a
function of time, as compared to the true clock errors. The clock
offsets estimates seem to converge asymptotically to the true
underlying states. The ionospheric parameter estimates are more
subject to ill-conditioning, as some parts of the TEC-screen are not
pierced by any projected station. However, plotting the TEC-screen
corresponding to the individual Legendre coefficients gives a good
qualitative match to the true TEC values (Fig. \ref{fig:SimsIon}).

\subsection{Pointing errors}
\label{sec:PointingWsrt}


One of the dominating calibration errors for interferometers using
dishes are the individual antenna pointing errors. They start to have
a significant effect even at moderate dynamic range, and can be severe
for non-symmetric primary beams with azimuthal dish
mounts. \citet{SB:pointing} and \citet{Smirnov2011qmc} have presented a
Physics-based calibration scheme to specifically solve for pointing
errors, using a least-squares minimization technique combined with a beam
model.


Here, we simulate a Westerbork Synthesis Radio Telescope (WSRT)
data-set. As in Sec. \ref{sec:IonClockSim}, we first define a measurement
equation (operator $\textbf{h}$). We only consider the direction dependent primary beam
effect, using the WSRT $\cos^3$ beam model:

\begin{alignat}{3}
\textbf{G}_{pt\nu}(\PVec)&:=&&\textbf{I}\\
\textbf{D}^d_{pt\nu}(\PVec)&:=&&\cos\left(\min\left\{65\ [\nu/10^9]\ r^d_{pt}(\PVec),1.0881\right\}\right)^3\textbf{I}\\
r^d_{pt}(\PVec) &=&& \sqrt{(\l^d-\delta\l_{pt}(\PVec))^2+(\m^d-\delta\m_{pt}(\PVec))^2}
\end{alignat}

\noindent where $\delta\l_{pt}$ and $\delta\m_{pt}$
are the operators unpacking the pointing errors values $\delta\l$ and
$\delta\m$ for antenna $p$ at time $t$. The
$\textbf{f}$ evolution operator is the same as in
Sec. \ref{sec:SimulOper}, with $\tau_i=5$ min
(Eq. \ref{eq:EvolOp}). We simulate a data-set containing $64$ channels
centered at $\sim 1.3$ GHz and channel width $\delta \nu=0.3$ MHz. The
sky model has $20$ sources with a total flux density of $\sim10$ Jy,
with a noise of $0.2$ Jy per visibility. The simulated pointing errors
have an initial global offset distributed as $\delta
\l_0\sim\delta\m_0\sim\mathcal{N}(0, 3)$ arcmin, and the pointing
offsets evolve periodically as $\delta
\l(t)=l_0+a_l\cos(2\pi t/\tau_l +\phi_l)$, with $a_l\sim\mathcal{N}(1,
0.3)$ arcmin, $\tau_l\sim\mathcal{N}(40, 10)$ min, and
$\phi_l$ uniformly distributed between $0$ and $2\pi$. The same scheme
is used to generate the evolution law for $\delta\m$).

Fig. \ref{fig:SimsWsrt} shows the comparison between the estimated
pointing errors are the true pointing errors for a few antennas. The filter's estimate
rapidly converges to the true pointing offset, and properly tracks its
state within the estimated uncertainty.


\section{Discussion and conclusion}
\label{sec:discussion}


\subsection{Overview: pros, cons and potential}
\label{sec:overview}


As discussed throughout this paper, it is important to obtain robust
algorithms that do not affect the scientific signal. 
In this paper, we have presented a method that aims at improving
robustness along the following lines:

\begin{enumerate}[(a)]
\item\label{enum:KF} 
The Kalman filter presented here is fundamentally
recursive, and information from the past is transferred along the
recursion, thereby constraining the expected location of the
underlying true state. This is fundamentally different from minimizing
a least square, and then smoothing or interpolating the solution -
especially since we can assume a physical measurement and evolutionary
model.
\item\label{enum:ME} 
Contrarily to the Jones-based algorithms that have to deal with
hundreds of degrees of freedom, our algorithm follow a Physics-based
approach (see in Sec. \ref{sec:PhysPeal} for other Physics-based
methods). It aims at estimating the true underlying {\it physical}
term, potentially describing the Jones matrices of individual effects
everywhere in the baseline-direction-frequency space. The very inner
structure of the Radio Interferometry Measurement Equation (RIME) can be used
to constrain the solutions. This feature allows us to take into
account much bigger data chunks. Typically, most effects have a very
stable frequency behaviour, and the data in the full instrumental
bandwidth can be simultaneously used at each recursion. This improves
conditioning.
\item\label{enum:Data} 
The measurement operator is non-linear, and combining (\ref{enum:KF}) with (\ref{enum:ME}) is made possible by using
a modern non-linear version of the Kalman filter, together with the
representation operator presented in Sec. \ref{sec:ReprSec}.
\item\label{enum:Tiko} Ill-conditioning can still be significant if
  effects are analytically degenerate to some degree. We can modify
  the measurement operator to take external prior information into
  account (see Sec. \ref{sec:ImplAMM}), and reject solutions that
  are considered to be non-physical. For example, this can allow the
  user to provide the filter with an expected ionospheric power
  spectrum of the Legendre coefficients.
\item\label{enum:Sigs} One of the benefits of using filters
  is that they produce a posterior covariance matrix on
  the estimated process state. The covariance estimate should be
  reliable assuming the non-linearities are not too severe.
\end{enumerate}

Given the large size of our inverse problem, and in order to make any
algorithm practical, optimizing the computational cost is of prime
importance. Using the Woodbury matrix
identity (Sec. \ref{sec:Woodbury}), we re-factor the Unscented
Kalman Filter steps to make the algorithm practical. Even for the
moderately large simulations described in
Sec. \ref{sec:IonClockSim}, a 4-core CPU is able to constrain
solutions faster than real time. The need to access the data of all
frequencies simultaneously represents some technical problems, as these
are distributed over different cluster nodes.

An important potential problem with Physics-based approaches is that the system
needs to be described analytically, while algorithms solving for the
effective Jones matrices do not use any assumptions about the physics
underlying the building of a visibility (apart from the sky model that
is assumed). This would cause problems in particular if the model
encapsulated in the operator $\textbf{h}$ misses physical ingredients
that {\it are} present in reality, and would probably drive biases in the estimates.
Furthermore, the Unscented Kalman Filter used and adapted to the context
of radio-interferometry in this paper deals with non-linearities only up to
a certain level. This means in practice that the process {\it a
  priori} covariance has a certain maximum size, for a given type of
non-linearities. 

\subsection{Conclusion}

The use of filters and similar methods can
potentially improve radio interferometric calibration. 
As
with least squares minimization techniques, our approach is guaranteed
to work only if non-linearities are not too severe in the neighborhood
of the true process state. Other algorithms dealing with
non-linearities are known to provide higher robustness, such as more
general particle filters, or Monte-Carlo Markov Chains. The later is
indeed guaranteed to provide a correct estimated posterior
distribution. However, most of these methods are expensive because of
the many predict steps that have to be computed, and this fact
could make them impractical, given the large size of our
problem. Recursive algorithms are well adapted to streaming
pre-calibration, and based on preliminary simulations, our algorithm
seems to be robust enough to solve for the sky term (positions, flux
densities, spectral indices, etc.) in a streaming way.

We have not yet demonstrated the efficiency of our algorithm with
real datasets essentially because of its complexity and novelty. Indeed, our
software needs to deal with a number of technical issues as well as more
fundamental problems. Specifically in the case of the newest
interferometers such as LOFAR, (i) we have to deal with large
quantities of distributed data, and the algorithm has to access all
frequencies simultaneously. Beyond these technical aspects, because we
solve for the underlying physical effects, (ii) we need to build
pertinent physical models for the various effects we are solving for,
such as ionosphere, or phased array beams. An application of this algorithm to
real datasets will be presented in a future paper.


\begin{acknowledgements}
I thank Ludwig Schwardt for helping me understand some important
aspects of Kalman filters. Those open-ended
discussions were very helpful to develope the framework presented in
this paper. Thanks to Trienko Grobler and Oleg Smirnov for giving
useful comments on the paper draft.
\end{acknowledgements}

\bibliographystyle{aa}
\bibliography{references}

\begin{thebibliography}{27}
\expandafter\ifx\csname natexlab\endcsname\relax\def\natexlab#1{#1}\fi

\bibitem[{Bhatnagar {et~al.}(2004)Bhatnagar, Cornwell, \& Golap}]{SB:pointing}
Bhatnagar, S., Cornwell, T.~J., \& Golap, K. 2004, EVLA Memo 84. Solving for
  the antenna based pointing errors, Tech. rep., NRAO

\bibitem[{{Bhatnagar} {et~al.}(2008){Bhatnagar}, {Cornwell}, {Golap}, \&
  {Uson}}]{Bhatnagar08}
{Bhatnagar}, S., {Cornwell}, T.~J., {Golap}, K., \& {Uson}, J.~M. 2008, \aap,
  487, 419

\bibitem[{{Bhatnagar} {et~al.}(2013){Bhatnagar}, {Rau}, \&
  {Golap}}]{Bhatnagar12}
{Bhatnagar}, S., {Rau}, U., \& {Golap}, K. 2013, \apj, 770, 91

\bibitem[{{Condon} {et~al.}(1998){Condon}, {Cotton}, {Greisen}, {Yin},
  {Perley}, {Taylor}, \& {Broderick}}]{Condon98}
{Condon}, J.~J., {Cotton}, W.~D., {Greisen}, E.~W., {et~al.} 1998, \aj, 115,
  1693

\bibitem[{{Cotton}(1995)}]{Cotton95}
{Cotton}, W.~D. 1995, in Astronomical Society of the Pacific Conference Series,
  Vol.~82, Very Long Baseline Interferometry and the VLBA, ed. J.~A. {Zensus},
  P.~J. {Diamond}, \& P.~J. {Napier}, 189

\bibitem[{{Ding} {et~al.}(2007){Ding}, {Wang}, {Rizos}, \&
  {Kinlyside}}]{Ding07}
{Ding}, W., {Wang}, J., {Rizos}, C., \& {Kinlyside}, D. 2007, Journal of
  Navigation, 60, 517

\bibitem[{{Hager}(1989)}]{Hager89}
{Hager}, W.~W. 1989, Society for Industrial and Applied Mathematics

\bibitem[{{Hamaker} {et~al.}(1996){Hamaker}, {Bregman}, \& {Sault}}]{Hamaker96}
{Hamaker}, J.~P., {Bregman}, J.~D., \& {Sault}, R.~J. 1996, \aaps, 117, 137

\bibitem[{{Henar}(2011)}]{Henar09}
{Henar}, F.~E. 2011, Master thesis, Institut f\"ur Biomedizinische Technik

\bibitem[{{Hiltunen} {et~al.}(2011){Hiltunen}, {S{\"a}rkk{\"a}}, {Nissil{\"a}},
  {Lajunen}, \& {Lampinen}}]{Hiltunen10}
{Hiltunen}, P., {S{\"a}rkk{\"a}}, S., {Nissil{\"a}}, I., {Lajunen}, A., \&
  {Lampinen}, J. 2011, Inverse Problems, 27, 025009

\bibitem[{{Intema} {et~al.}(2009){Intema}, {van der Tol}, {Cotton}, {Cohen},
  {van Bemmel}, \& {R{\"o}ttgering}}]{Intema09}
{Intema}, H.~T., {van der Tol}, S., {Cotton}, W.~D., {et~al.} 2009, \aap, 501,
  1185

\bibitem[{{Julier} \& {Uhlmann}(1997)}]{Julier97}
{Julier}, S.~J. \& {Uhlmann}, J.~K. 1997, in Society of Photo-Optical
  Instrumentation Engineers (SPIE) Conference Series, Vol. 3068, Signal
  Processing, Sensor Fusion, and Target Recognition VI, ed. I.~{Kadar},
  182--193

\bibitem[{{Junklewitz} {et~al.}(2014){Junklewitz}, {Bell}, \&
  {En{\ss}lin}}]{Junklewitz14}
{Junklewitz}, H., {Bell}, M.~A., \& {En{\ss}lin}, T. 2014, ArXiv e-prints

\bibitem[{Kalman(1960)}]{Kalman60}
Kalman, R.~E. 1960

\bibitem[{{Kazemi} {et~al.}(2011){Kazemi}, {Yatawatta}, {Zaroubi},
  {Lampropoulos}, {de Bruyn}, {Koopmans}, \& {Noordam}}]{Kazemi11}
{Kazemi}, S., {Yatawatta}, S., {Zaroubi}, S., {et~al.} 2011, \mnras, 414, 1656

\bibitem[{{Mandel}(2006)}]{Mandel06}
{Mandel}, J. 2006, in CCM Report 231, University of Colorado at Denver and
  Health Sciences Center

\bibitem[{{McEwen} \& {Wiaux}(2011)}]{McEwen11}
{McEwen}, J.~D. \& {Wiaux}, Y. 2011, ArXiv e-prints

\bibitem[{{Noordam} \& {Smirnov}(2010)}]{Noordam10}
{Noordam}, J.~E. \& {Smirnov}, O.~M. 2010, \aap, 524, A61

\bibitem[{{Rau} \& {Cornwell}(2011)}]{Rau11}
{Rau}, U. \& {Cornwell}, T.~J. 2011, \aap, 532, A71

\bibitem[{{Smirnov}(2011)}]{Oleg11}
{Smirnov}, O.~M. 2011, \aap, 527, A106

\bibitem[{Smirnov(2011)}]{Smirnov2011qmc}
Smirnov, O.~M. 2011, presentation at CALIM2011 conference

\bibitem[{{Tasse} {et~al.}(2013){Tasse}, {van der Tol}, {van Zwieten}, {van
  Diepen}, \& {Bhatnagar}}]{Tasse13}
{Tasse}, C., {van der Tol}, S., {van Zwieten}, J., {van Diepen}, G., \&
  {Bhatnagar}, S. 2013, \aap, 553, A105

\bibitem[{{Walker}(1999)}]{Walker99}
{Walker}, R.~C. 1999, in Astronomical Society of the Pacific Conference Series,
  Vol. 180, Synthesis Imaging in Radio Astronomy II, ed. G.~B. {Taylor}, C.~L.
  {Carilli}, \& R.~A. {Perley}, 433

\bibitem[{{Wan} \& {van der Merwe}(2000)}]{Wan00}
{Wan}, E.~A. \& {van der Merwe}, R. 2000, The Unscented Kalman Filter for
  Nonlinear Estimation

\bibitem[{{Yatawatta}(2013)}]{Yatawatta13}
{Yatawatta}, S. 2013, Experimental Astronomy, 35, 469

\bibitem[{{Yatawatta} {et~al.}(2008){Yatawatta}, {Zaroubi}, {de Bruyn},
  {Koopmans}, \& {Noordam}}]{Yatawatta08}
{Yatawatta}, S., {Zaroubi}, S., {de Bruyn}, G., {Koopmans}, L., \& {Noordam},
  J. 2008, ArXiv e-prints

\bibitem[{{Yavuz}(2007)}]{Yavuz07}
{Yavuz}, E., A.~F. A. O. E. C.~B. 2007, 13th, European Signal Processing
  Conference

\end{thebibliography}


\begin{appendix}
\section{One-dimensional images properties}
\label{sec:1Ddetail}

In this section we discuss in more detail the one-dimensional image
representations introduced in Sec. \ref{sec:ReprSec}. 
Aiming at being as conservative
as possible, but still working in the image domain, we define
$\OpIm1D$ to be the operator that builds one 1D image per baseline $(pq)$,
and polarization $i$ as:

\begin{alignat}{3}
\nonumber\widetilde{\textbf{y}}_{(pq),i}&=\OpIm1D(\VisVec_{(pq),i})\\
\nonumber              &:=\int_{-\infty}^{\infty} \VisVec_{(pq),i}(\nu)\ 
\exp\left(
2\pi i \frac{\nu}{\text{c}} \textbf{b}_{pq}^T\textbf{s}
\right) d\nu\\
              &=\int_{-\infty}^{\infty} \textbf{V}_{(pq),i}(\nu)\ \mathrm{rect}\left(
\frac{\nu-\nu_m}{\Delta \nu} \right)
\exp\left(
2\pi i \frac{\nu}{\text{c}} \textbf{b}_{pq}^T\textbf{s}
\right) d\nu
\end{alignat}

\begin{alignat}{3}
\text{with }\textbf{V}_{(pq),i}&=\sum_{d=1}^{n_d}\text{S}^d_{i}\exp\left(
-2\pi i \frac{\nu}{\text{c}} \textbf{b}_{pq}^T\textbf{s}_d
\right) d\nu
\end{alignat}

\noindent where $c$ is the speed of light, $\mathrm{rect}$ is the
rectangular function, $\nu_m=(\nu_0+\nu_1)/2$, $\Delta
\nu=\nu_1-\nu_0$ with $\nu_0$ and $\nu_1$ the minimum and maximum
available frequencies. 
In order to align the $u$-coordinate with the frequency extent of the
baseline, we rotate $\textbf{b}_{pq}$ and $\textbf{s}$ and
$\textbf{s}_d$ with $3\times3$ rotation matrix $\textbf{U}_{\theta\phi}$
such that:

\begin{alignat}{2}
\textbf{b}_{\theta\phi}&=\textbf{U}_{\theta\phi}
\left [
\begin {array}{c}
u_{pq}\\ v_{pq}\\ w_{pq}
\end {array}
\right ]
&&=
\left [
\begin {array}{c}
u_{pq}'\\ 0 \\ 0
\end {array}
\right ]\\
\textbf{s}_{\theta\phi}&=\textbf{U}_{\theta\phi}
\left [
\begin {array}{c}
l\\ m\\n-1
\end {array}
\right ]
&&=
\left [
\begin {array}{c}
l'\\ m'\\n'
\end {array}
\right ]
\end{alignat}

\noindent where $(l, m, n)$
are the image plane coordinates, $(u_{pq}, v_{pq})$ are the
uv-coordinate of baseline $(pq)$, $(\nu_0,\nu_1)$ are the lower and higher frequencies
values of the interferometer's bandpass. It is unitary so
$\textbf{b}^T\textbf{s}=\textbf{b}_{\theta\phi}^T\textbf{s}_{\theta\phi}$. 
We can write the
complex 1-dimensional image as:

\begin{alignat}{3}
\widetilde{\textbf{y}}_{(pq),i}&=\sum_{d=1}^{n_d}\text{S}^d_{i}\delta({l'_d})*\text{PSF}_{1D}
\end{alignat}

\noindent where $*$ is the convolution product, $\text{S}_{d,i}$ is the
apparent flux of the source in direction $d$ for polarization $i$.

More intuitively, this means that $\widetilde{\textbf{y}}_{(pq),i}$ is obtained by projecting the sky on the
baseline, and convolving it with the 1-dimensional PSF of the given baseline. 
The term $\text{PSF}_{1D}$ is obtained by computing the inverse Fourier transform of the
uv-domain sampling function:

\begin{alignat}{3}
\text{PSF}_{1D}(l')&=\mathcal{F}^{-1}\left\{
\mathrm{rect}\left(
\frac{\nu-\nu_m}{\Delta \nu} \right)
\right\}\\
\label{eq:PSF1D}
&=\textrm{sinc}\left(u'\Delta \nu l'/c\right)\exp{\left(2\pi i u'\nu_ml'/c\right)}
\end{alignat}

\noindent where $\textrm{sinc}$ is the cardinal sine
function. We can see in Eq. \ref{eq:PSF1D} that $\text{PSF}_{1D}$ contains both low and high spacial
frequency terms ($u'\Delta \nu /c$ and $u'\nu_m/c$
respectively, whose ratio equals the fractional bandwidth).

Therefore, while $\RepImRe$ still contains the high frequency fringe,
taking the complex norm using $\RepImAbs$ eliminates the high spacial
frequency term, and $\text{PSF}_{1D}$ under $\RepImAbs$ is
smoother ($\RepImRe$ extracts the envelope of $\text{PSF}_{1D}$). This intuitively explains why the $\RepImAbs$ representation
seems to provide a better match between the true and the
$\sigma$-points statistics. As the clock and ionospheric displacements
mostly amount to apparent shifts in source locations, smoothness of
$\RepImAbs$ provides stability in the $\sigma$-points statistics in
the data domain. 



Applying any of the
representation operators presented above modifies the properties
of the input data covariance matrix $\MR$ (Eq. \ref{eq:KF_Sk},
\ref{eq:UKF_Pzzh}, and \ref{eq:UKF_Pzzh_S}). Assuming the
noise in the visibilities is uncorrelated, $\MR$ is diagonal
matrix. Assuming the noise is not baseline or frequency dependent we have
$\MR=\sigma^2\ \bm{\mathcal{I}}$, for $\RepUV$ and $\RepUVRe$,
and $\MR=(\sigma^2/n_\nu)\ \bm{\mathcal{I}}$ for $\RepImRe$. The
statistics of $\RepImAbs$ are non-Gaussian since it is the norm of a
complex number. The random variable is in that case $X=\sigma^{-1}\sqrt{n_\nu}\sqrt{
  (Re^2+Im^2)}$ which follows a non-central $\chi$-distribution with
2-degrees of freedom, and mean and covariance:

\begin{alignat}{3}
\mu_X&=\sqrt{\frac{\pi}{2}}L_{1/2}^{0}\left\{\frac{-\lambda^2}{2}\right\}\\
\sigma_X^2&=2+\lambda^2-\mu_X^2\\
\text{with }\lambda&=\sqrt{n_\nu}\sigma^{-1}\sqrt{\mu_{Re}^2+\mu_{Im}^2}
\end{alignat}

\noindent where $L_{1/2}^{0}$ is a generalized Laguerre polynomial.

\end{appendix}

\end{document}